\documentclass[aps,prd,superscriptaddress,showpacs,preprintnumbers]{revtex4}
\usepackage{graphicx}
\usepackage{mathrsfs}
\usepackage{amssymb}
\usepackage{bm}
\usepackage{hyperref}

\usepackage{color}

\newcommand{\ba}{\begin{eqnarray}}
\newcommand{\ea}{\end{eqnarray}}

\newcommand{\eps}{\epsilon}
\newcommand{\ii}{i}
\newcommand{\g}{\gamma}

\usepackage{graphics}
\usepackage{bm}
\usepackage{amsmath}
\usepackage{amssymb}

\newcommand{\nslash}{\kern 0.2 em n\kern -0.50em /}
\newcommand{\kslash}{\kern 0.2 em k\kern -0.45em /}
\newcommand{\pslash}{\kern 0.2 em p\kern -0.50em /}
\newcommand{\Sslash}{\kern 0.2 em S\kern -0.50em /}
\newcommand{\Pslash}{\kern 0.2 em P\kern -0.50em /}
\newcommand{\Dslash}{\kern 0.2 em D\kern -0.65em /\kern 0.15em}
\newcommand{\slim}{\mskip 1.5mu}
\newcommand{\cdott}{{\mskip -1.5mu} \cdot {\mskip -1.5mu}}

\newcommand{\pp}{\,\!^\perp}
\begin{document}

\title{Transverse momentum dependent twist-three result for polarized Drell-Yan processes}

\author{Zhun Lu}
\affiliation{Department of Physics, Southeast University, Nanjing
211189, China}

\author{Ivan Schmidt}
\affiliation{Departamento de F\'\i sica, y
Centro Cient\'ifico-Tecnol\'ogico de Valpara\'iso, Universidad T\'ecnica Federico Santa Mar\'\i a,
Casilla 110-V, Valpara\'\i so, Chile}

\begin{abstract}
We study the polarized Drell-Yan processes from the collision of two spin-$1/2$ hadrons
at order $1/Q$ based on the framework of transverse momentum dependent
factorization.
We give the complete twist-three results of total sixteen independent structure
functions in terms of twist-two and twist-three transverse momentum dependent
distribution functions.

\end{abstract}

\pacs{12.38.Bx, 12.39.St, 13.85.Qk}

\maketitle

\section{introduction}

Polarized Drell-Yan processes are a promising ground for the study of
nucleon structure~\cite{Ralston:1979ys,Jaffe:1991kp}.
If the transverse momentum of the lepton pair is detected, polarized Drell-Yan processes
can be used to probe~\cite{tm94,Boer1999} various transverse momentum dependent (TMD) parton distributions,
which have received considerable interest recently (e.g., see \cite{bdr,Boer:2011fh,Barone:2010ef,
D'Alesio:2007jt} and reference therein).
Compared with semi-inclusive deep inelastic scattering (SIDIS), which has been used
intensively to study TMD distributions in the last decade~\cite{
Airapetian:1999tv,
smc,Airapetian:2001eg,Avakian:2003pk,
Airapetian:2004tw,compass,hermes05,compass06,
Mkrtchyan:2007sr,2009ti,Alekseev:2010rw,Qian:2011py},
Drell-Yan reactions have the feature that only parton distributions are involved, that is,
there is no hadron detected in the final state.

Theoretically, the TMD factorization for Drell-Yan process at low transverse momentum has
been established, and also the complete leading-twist Drell-Yan structure functions for
spin-$1/2$ hadrons beams have been given in Ref.~\cite{Arnold2009}.
Previous studies on Drell-Yan processes based on TMD framework mainly focused on
the contribution at leading power of $1/Q$, where $Q$ is the invariant mass of
the lepton pair.
This motivate us to present a full expression for the Drell-Yan process at
twist-three level with dilepton transverse momentum kept unintegrated, which is
the main goal of this paper.
We note that at order $1/Q$, results contributed by the (transverse momentum) integrated distributions
$f_T(x)$, $g_T(x)$, $h_L(x)$ and $h(x)$ have been worked out in
Refs.~\cite{Tangerman:1994bb,Boer:1997bw}.
We will consider both polarized and unpolarized scattering of
spin-$1/2$ hadron beams.
We find that at order $1/Q$ there are sixteen transverse momentum dependent structure
functions for the Drell-Yan process, which can be expressed as a convolution of
twist-two and twist-three TMD distributions.

Experimentally, a number of polarized Drell-Yan programs have been proposed
at several facilities~\cite{PAX_05,RHIC_08,COMPASSpro,Goto:2010zz,Goto:2011zz,
Meshkov:2011zz,Liu:2010kb,Vasiliev:2007nz}
and some of them could be realized in the near future
to provide the first polarized data.
The twist-three contributions can be potential experimental observables and
may be accessible in certain kinematical regions.
The interest on the twist-three contributions also comes from the fact
that they are related to the quark-gluon correlation inside the nucleon~\cite{Jaffe:1989xx,Burkardt:2008ps},
which is still not understood yet.

We need to emphasize that the approach in this paper is based on the assumption that
the framework of TMD factorization is valid at order $1/Q$.
The same approach has been applied in Ref.~\cite{Bacchetta:2004zf,Bacchetta:2006tn} to calculate the complete
leading-twist and subleading-twist observables in SIDIS (for the production of spin-0 hadrons),
where ten twist-three TMD structure functions have been found.
Therefore our twist-three results should not be compared with the twist-three mechanism
~\cite{Efremov:1981sh,Efremov:1984ip,Qiu:1991pp,Qiu:1998ia,Koike:2007dg,Ji:2006vf}
in {\it collinear factorization} that has been applied to study the single-spin asymmetries
in Drell-Yan processes~\cite{Ji:2006vf}.

The remaining content of the paper is organized as follows.
In Section.~II we introduce the formalism needed in the construction of
TMD twist-three observables.
In Section. III we present the complete expressions for the differential cross section of
Drell-Yan process with dilepton transverse momentum unintegrated at order $1/Q$.
We summarize the paper in Section. IV.

\section{Formalism and kinematical settings}

The process we study is
\begin{eqnarray}
h_1 (P_1,S_1) + h_2 (P_2,S_2)  \to \ell(l) + \bar{\ell}(l^\prime) +X. \label{hhdy}
\end{eqnarray}
Here we consider only the electromagnetic interaction. The notations $P_i$ and $S_i$ are the four-momenta and spins of the hadron beams which can be decomposed as
\begin{eqnarray}
P_1^\mu & = & P_1^+\,n_+^\mu
+ {M_1^2 \over P_1^+}\,n_-^\mu,\\
P_2^\mu & = & {M_2^2 \over P_2^-}\,n_+^\mu
+ P_2^- \,n_-^\mu,
\\[2 mm]
S_1 & = & \frac{\lambda_1 P_1^+}{ M_1 }\,n_+^\mu
-\frac{ \lambda_1 M_1}{ P_1^+ }\,n_-^\mu + S_{1T}^\mu
,\\[2 mm]
S_2 & = & \frac{\lambda_2 P_2^-}{M_2 }\,n_-^\mu
-\frac{\lambda_2 M_2}{P_2^-}\,n_+^\mu + S_{2T}^\mu,
\end{eqnarray}
where $n_+ $ and $n_-$ are two light-like vectors expressed in the
light-cone coordinates, in which an arbitrary four-vector $a$ is
written as $\{a^-,a^+, \boldsymbol{a}_T\}$, with $a^{\pm}=(a^0 \pm
a^3)/\sqrt{2}$ and $\boldsymbol{a}_T =(a^1,a^2)$.
Then one can define following transverse tensors
\begin{eqnarray}
g_T^{\mu\nu} =g^{\mu\nu} - n_+^\mu n_-^\nu - n_-^\mu n_+^\nu,~~~\epsilon_T^{\mu\nu} =\epsilon^{\mu\nu\rho\sigma}
n_{+\rho} n_{-\sigma}.
\end{eqnarray}
We will apply the parton model to study the Drell-Yan pair production. In this model, the leading contribution is from the annihilation of the quark and antiquark from each proton: $q(k_1) q(k_2)\rightarrow \gamma^* \rightarrow  \ell  \bar{\ell} $.
The momenta of the quark, antiquark and the virtual photon can be decomposed as
\ba
k_1 & = & x_1\,P_1^+ \,n_+
+ \frac{(k_1^2 + \bm{k}_{1T}^2)}{x_1 P_1^+}\,n_- + k_{1T}
, \\[2 mm]
k_2 & = & x_2\,P_2^-\,n_-
+ \frac{ (k_2^2 + \bm{k}_{2T}^2)}{x_2 P_2^-}\,n_+ + k_{2T}
,\\[2 mm]
q  & = &\frac{ Q}{\sqrt{2}}\,n_+
+ \frac{ Q}{\sqrt{2}}\,n_-^\mu
+ q_T,
\ea
where $Q^2=q^2$, and we limit our study to the region $Q_T^2 = \bm q_T^2 = - q_T^2 \ll Q^2 $, and thus the intrinsic transverse momenta of quarks play significant role.

\begin{figure}
\begin{center}
\scalebox{1.2}{\includegraphics{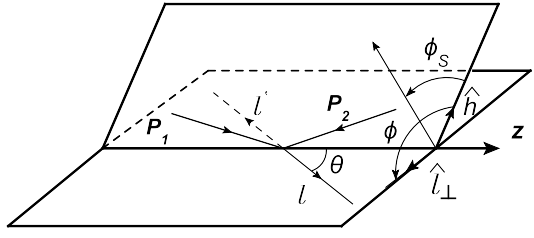}}\caption{\small  Angular
definitions of polarized Drell-Yan process in the lepton pair
center of mass frame.}\label{fig:dyframe}
\end{center}
\end{figure}

The angular distribution of the Drell-Yan cross section is usually expressed in the dilepton rest frame (Fig.\ \ref{fig:dyframe}), which can be defined by introducing the following normalized vectors~\cite{Boer:1997bw}:
\begin{eqnarray}
& & \hat t = q/Q,\\
& & \hat z =  (1-c) \frac{2x_1}{Q}
\tilde{P_1}- c \frac{2x_2}{Q} \tilde{P_2},\\
& & \hat h = q_T/Q_T = (q-x_1\, P_1 -x_2\, P_2)/Q_T,
\end{eqnarray}
where $\tilde{P_i} = P_i-q/(2 x_i)$.
The parameter $c$ represents the degree of freedom to distribute the transverse momentum between
$P_1$ and $P_2$.
The cases $c=0,\, 1/2$ and $1$ correspond to the Gottfried-Jackson frame~\cite{Gottfried:1964nx}, the Collins-Soper frame~\cite{CS_frame} and the $u$-channel frame, respectively.
Using the normalized vectors $\hat t$ and $\hat z$ one can construct
the perpendicular tensors as
\begin{eqnarray}
g_\perp^{\mu\nu} =g^{\mu\nu} - \hat t^\mu \hat t^\nu + \hat z^\mu \hat z^\nu,~~~\epsilon_\perp^{\mu\nu} =-\epsilon^{\mu\nu\rho\sigma}
\hat t_{\rho} \hat z_{\sigma}.
\end{eqnarray}
As shown in Fig.~\ref{fig:dyframe}, the azimuthal angles of the dilepton and of the transverse spin of the hadrons with respect to the hadron plane can defined as~\cite{Bacchetta:2004jz}
\begin{eqnarray}
& & \cos\phi = -g_\perp^{\mu\nu} \hat h_\mu \hat l_{\perp\nu} , ~~~~
\sin \phi =\epsilon_{\perp}^{\mu\nu} \hat h_\mu  \hat l_{\perp\nu}, \label{eq:phi_def} \\
& & \cos\phi_S = {-g_\perp^{\mu\nu} \hat h_\mu S_{\perp\nu}\over \sqrt{- S_\perp^2}},~~  \sin \phi_S ={ \epsilon_{\perp}^{\mu\nu} \hat h_\mu   S_{\perp\nu}\over \sqrt{-S_{\perp}^2}},\label{eq:phi_S_def}
\end{eqnarray}
where $\hat l_\perp^\mu = g_\perp^{\mu\nu} l_\nu/\sqrt{-g_\perp^{\mu\nu}l_\mu l_\nu}$£¬ and $S_\perp^\mu = g_\perp^{\mu\nu} S_\nu$.
We note that the definition in Eqs.(\ref{eq:phi_def}) and (\ref{eq:phi_S_def}) is the same as the one used in Ref.~\cite{Arnold2009}.
The two light-like vectors can be expanded as linear combinations of $\hat t$, $\hat z$ and the perpendicular vector $\hat h$
\begin{eqnarray}
n_+^\mu & = &
\frac{1}{\sqrt{2}} \left[ \hat t^\mu + \hat z^\mu
-2c\,\frac{Q_T^{}}{Q} \hat h \right],
\label{transverse1} \\
n_-^\mu & = &
\frac{1}{\sqrt{2}} \left[ \hat t^\mu - \hat z^\mu
- 2(1-c)\,\frac{Q_T^{}}{Q}\,\hat h^\mu \right].
\label{transverse2}
\end{eqnarray}
Thus the transverse tensor and the perpendicular tensor are related by
\ba
g_T^{\mu\nu} =g_\perp^{\mu\nu} +{Q_T\over Q} \hat t^{\{\mu} \hat h^{\nu\}} +(1-2c){Q_T\over Q} \hat z^{\{\mu} \hat h^{\nu\}} + \mathcal{O} (1/Q^2). \label{gtgperp}
\ea
where the symmetrization of indices is used.
One can see that the differences are of order $1/ Q$.
As we study the twist-three contribution, we need to keep
track of these differences.

In the rest frame of the dilepton, one can express the differential cross section of the Drell-Yan process as
\begin{equation} \label{eq:dycs}
\frac{d\sigma}{d^4 q \, d\Omega}
= \frac{\alpha_{em}^2}{2 s \, q^4} \, L_{\mu\nu} W^{\mu\nu} \,,
\end{equation}
where $d\Omega=d\cos\theta d
\phi$ is the solid angle of the lepton $\ell$.
The notation $L^{\mu\nu}$ denotes the lepton tensor which has the following form~\cite{Tangerman:1994bb}:
\begin{eqnarray}
L^{\mu \nu} & = &Q^2 \Biggl[
- \left( {1+\cos^2\theta \over 2} \right) g_\perp^{\mu \nu}
+ \sin^2\theta \hat z^\mu \hat z^\nu
\nonumber \\ && \qquad
-\sin^2\theta\left(  \hat l^\mu_\perp  \hat l^\nu_\perp +\frac{1}{2}\,g_\perp^{\mu \nu}
\right)
+ \sin 2 \theta \,\,\hat z_{\rule{0mm}{3mm}}^{\{ \mu}\hat l_\perp^{\,\nu \}}
\Biggr],
\end{eqnarray}
Here we have ignored the lepton masses and their polarization.

The hadronic tensor $W^{\mu\nu}$ can be expressed as~\cite{Boer:2003cm}
\begin{align}
W^{\mu\nu} & =\frac{1}{3}
\sum_a   e_a^2 \; \int
d^2 \bm{k}_{1T}\ d^2 \bm{k}_{2T}^{}\
\delta^2(\bm{k}_{1T} + \bm{k}_{2T} - \bm{q}_T^{}) \;
\textrm{Tr} \biggl\{
  \Phi^a (x,k_{1T}) \gamma^\mu \overline\Phi^{\,a} (x_2,k_{2T}) \gamma^\nu
\\
&+ \frac{1}{Q\sqrt{2}} \biggl[
  \gamma^\alpha \nslash_+ \gamma^\nu \slim
  \tilde{\Phi}^a_{A\slim \alpha}(x,k_{1T})\slim \gamma^\mu {\overline \Phi}^a(x,k_{2T})
+ \gamma^\mu \nslash_+ \gamma^\alpha\overline{\Phi}(x,k_{2T}) \gamma^\mu
  \tilde{\Phi}^a_{A\slim \alpha} (x,k_{1T})
   \slim\biggr] \nonumber\\
&-\frac{1}{Q\sqrt{2}} \biggl[
  \gamma^\nu \nslash_- \gamma^\alpha  \Phi^a(x,k_{1T})\slim \gamma^\mu
  \tilde{\overline{\Phi}}\,\!^a_{A\slim \alpha}(x,k_{1T})\slim
+ \gamma^\alpha \nslash_- \gamma^\mu
  \tilde{\overline{\Phi}}\,\!^a_{A\slim \alpha} (x,k_{2T})\slim \gamma^\nu
  \Phi^a (x,k_{1T})
   \slim\biggr] \biggr\} + \left(\begin{array}{c}
q\leftrightarrow -q \\ \mu \leftrightarrow \nu
\end{array} \right),\label{htensor}
\end{align}
where the factor $1/3$ takes into account color average, $a$ is the flavor index
and $e_a$ denotes the charge for flavor $a$.
The first line in the curly brackets in Eq.~(\ref{htensor}) comes from the diagram
without additional gluon connecting to the soft parts: the gauge-invariant TMD
dependent quark-quark correlation function $\Phi(x,k_{1T})$ and antiquark-antiquark
correlation function $\overline{\Phi}(x,k_{2T})$.
The second and third lines in the curly brackets in Eq.~(\ref{htensor}) correspond to
the diagrams involving one gluon which connects to one of the two soft parts, represented
by the quark-gluon-quark correlator $\Phi^a_{A\alpha}(x,k_{1T})$ or the
antiquark-gluon-antiquark correlator $\tilde{\overline \Phi}\,\!^a_{A\alpha}(x,k_{1T})$, with $\alpha$
restricted to be the transverse index.
Up to twist-three level, the TMD correlator $\Phi(x,k_T)$ can be
parameterized as~\cite{Goeke:2005hb,Bacchetta:2006tn}
\begin{align}
\Phi(x,k_T) &= \frac{1}{2}\, \biggl\{
f_1 \nslash_+
- {f_{1T}^\perp}\, \frac{\eps_T^{\rho \sigma} k_{T\rho}^{}\slim
  S_{T\sigma}^{}}{M} \, \nslash_+
+ g_{1s} \gamma_5\nslash_+
+h_{1T}\,\frac{\bigl[\Sslash_T, \nslash_+ \bigr]\gamma_5}{2}
+ h_{1s}^\perp \,\frac{\bigl[\kslash_T, \nslash_+ \bigr]\gamma_5}{2 M}
+\ii \, {h_1^\perp} \frac{ \bigl[\kslash_T, \nslash_+ \bigr]}{2M}
\biggr\}
\nonumber \\[0.2em] & \quad
+ \frac{M}{2 P^+}\,\biggl\{
e
- \ii\,{e_s} \,\gamma_5
- {e_{T}^\perp}\, \frac{\eps_T^{\rho \sigma} k_{T\rho}^{}\slim
  S_{T\sigma}^{}}{M}
+ f^\perp\, \frac{\kslash_T}{M}
- {f_T'}\,\epsilon_T^{\rho\sigma} \gamma_\rho^{}\slim S_{T \sigma}^{}
- {f_s^{\perp}}\,\frac{\eps_T^{\rho \sigma}
    \g_{\rho}^{}\slim k_{T \sigma}^{}}{M}+ g_T'\, \gamma_5\Sslash_T
\nonumber \\ & \quad \qquad \qquad
+ g_s^{\perp} \gamma_5 \frac{\kslash_T}{M}
- {g^\perp} \g_5\,\frac{\eps_T^{\rho \sigma}
    \g_{\rho}^{}\slim k_{T \sigma}^{}}{M}
+ h_s\,\frac{[\nslash_+, \nslash_-]\gamma_5}{2}
+ h_T^{\perp}\,\frac{\bigl[\Sslash_T, \kslash_T \bigr]\gamma_5}{2 M}
+ \ii \, {h} \frac{ \bigl[\nslash_+, \nslash_- \bigr]}{2}
\biggr\} .
\label{eq:phi}
\end{align}
The distribution functions on the right hand side (r.h.s.) of (\ref{eq:phi}) depend on $x$ and $k_T^2$,
except for the functions with subscript $s$, where the following
shorthand notation has been used~\cite{Mulders:1995dh}
\begin{equation}
g_{1s}(x, k_T^2)
= S_L\,g_{1L}(x, k_T^2) - \frac{k_T \cdott S_T}{M}\,g_{1T}(x, k_T^2)
\label{eq:sh}
\end{equation}
and so on for the other functions. The functions with subscript ``1" are the leading twist
distributions which have probability interpretations. The other sixteen functions are
twist-tree distributions. The calculations for eight T-even twist-three distributions has been carried in the diquark spectator model~\cite{Jakob:1997wg} and in the bag model~\cite{Avakian:2010br}.
There is also attempt to calculate naive-T-odd twist-three distributions~\cite{Gamberg:2006ru},
for which the light-cone divergence emerges.

The decomposition of antiquark correlator $\overline \Phi(x,k_T)$ can be achieved by the replacements
$x_1\rightarrow x_2, ~n_+\leftrightarrow n_-, ~\epsilon_T\rightarrow -\epsilon_T $, and by the relations $\overline{\Phi}^{[\Gamma]}=
 \overline{\Phi}^{c[\Gamma]}  $ for $\Gamma=\gamma_\mu, i\sigma_{\mu\nu}\gamma_5,i\gamma_5$ and $\overline{\Phi}^{[\Gamma]}= -
 \overline{\Phi}^{c[\Gamma]}  $ for $\bm{1}, i\gamma_\mu\gamma_5$~\cite{Mulders:1995dh}, where $c$ denotes the charge conjugation operation.

The quark-gluon-quark correlator $\Phi^a_{A\alpha}(x,k_{T})$ can be decomposed
as~\cite{Bacchetta:2006tn}
\begin{align}
\tilde\Phi_A^{\alpha}(x,k_T) =
 & \frac{x M}{2}\,
\biggl\{
\Bigl[
\bigl(\tilde{f}^\perp-\ii\slim \tilde{g}^{\perp} \bigr)
        \frac{k_{T \rho}^{}}{M}
-\bigl(\tilde{f}_T'+ \ii\slim \tilde{g}_T'\bigr)
     \,\epsilon_{T \rho\sigma}^{}\slim S_{T}^{\sigma}
-\bigl(\tilde{f}_s^{\perp}+\ii\,\tilde{g}_s^{\perp}\bigr)
     \frac{\epsilon_{T \rho \sigma}^{}\slim k_{T}^{\sigma}}{M}
 \slim\Bigl]
\bigl(g_T^{\alpha \rho} - i \epsilon_T^{\alpha\rho} \gamma_5\bigr)
\nonumber \\[0.2em]
 & \quad\!
-\bigl(\tilde{h}_s + \ii\,\tilde{e}_s\bigr)
        \gamma_T^{\alpha}\,\gamma_5
+\Bigl[\bigl(\tilde{h} + \ii\,\tilde{e}\bigr)
  +\bigl( \tilde{h}_T^{\perp}
        - \ii\,\tilde{e}_T^{\perp}\bigr)\slim
   \frac{\eps_T^{\rho \sigma} k_{T\rho}^{}\slim S_{T\sigma}^{}}{M}
 \slim\Bigr]
  \ii \gamma_T^{\alpha}
+ \ldots \bigl(g_T^{\alpha \rho}
               + \ii \epsilon_T^{\alpha\rho} \gamma_5\bigr)
\biggr\} \frac{\nslash_+}{2}\, ,
\label{eq:phiAalpha}
\end{align}
where the index $\alpha$ is restricted to be transverse.
The functions on the r.h.s\ with tilde are interaction-dependent twist-three
distributions.
They depend on $x$ and $k_T^2$, except for the functions with subscript $s$, which are
defined as in Eq.~(\ref{eq:sh}).
The last term inside the curly brackets will not be used in our following calculation
and can be omitted here. In the Wandzura-Wilczek approximation~\cite{Wandzura:1977qf}, the interaction-dependent distributions
are set to be zero. However, quantitative analysis~\cite{Accardi:2009au} on the $g_2$ structure function shows that the violation of
the Wandzura-Wilczek approximation is sizable.

According to the equation of motion for the quark field, relations between twist-two distributions
and twist-three distributions can be established~\cite{Mulders:1995dh}. A full list of these relations
can be found in Ref.~\cite{Bacchetta:2006tn}. Here we only quote the ones which will be used in our later
calculations:
\begin{align}
x f^{\perp} &=x  \tilde{f}^{\perp}+ f_{1},
\phantom{\frac{m^2}{M}}\label{eq:fperptilde}
\\
x g_T' &= x \tilde{g}_T'+\frac{m}{M}\,h_{1T},
\\
x g_T^{\perp}&= x  \tilde{g}_T^{\perp}+ g_{1T}
 +\frac{m}{M}\,h_{1T}^{\perp},
\\
x g_T &= x \tilde{g}_T -\frac{p_T^2}{2 M^2}\,  g_{1T}+\frac{m}{
  M}\,h_{1},
\\
x g_L^{\perp} &= x  \tilde{g}_L^{\perp} + g_{1L}+\frac{m}{
  M}\,h_{1L}^{\perp},
\\
x h_L &= x \tilde{h}_L +\frac{p_T^2}{M^2}\, h_{1L}^{\perp}
 +\frac{m}{M}\,g_{1L},
\\
x h_T &= x \tilde{h}_T - h_{1} +\frac{p_T^2}{2 M^2}\, h_{1T}^{\perp}
 +\frac{m}{M}\,g_{1T},
\\
x  h_T^{\perp} &= x \tilde{h}_T^{\perp}+h_{1}
 +\frac{p_T^2}{2 M^2}\, h_{1T}^{\perp},
\\
x f_T'  &= x  \tilde{f}_T' + \frac{p_T^2}{M^2}\, f_{1T}^{\perp},
\displaybreak[0]\\
x f_T^{\perp} &= x \tilde{f}_T^{\perp} +  f_{1T}^{\perp},
\phantom{\frac{p_T^2}{M^2}}
\\
x f_T &= x \tilde{f}_T +\frac{p_T^2}{2 M^2}\,  f_{1T}^{\perp},
\\
x f_L^{\perp}&= x  \tilde{f}_L^{\perp},
\phantom{\frac{p_T^2}{M^2}}
\\
x g^{\perp} &=  x  \tilde{g}^{\perp}+\frac{m}{M}\,h_{1}^{\perp},
\phantom{\frac{p_T^2}{M^2}}
\\
x h  &=x  \tilde{h} + \frac{p_T^2}{M^2}\, h_1^{\perp}. \label{eq:htilde}
\end{align}
We note that the Lorentz invariance relations~\cite{Tangerman:1994bb,Mulders:1995dh,Boer:1997nt,Metz:2008ib}  for integrated twist-tree distributions and their violations~\cite{Kundu:2001pk,Schlegel:2004rg,Goeke:2003az} have been discussed in literature.

\section{Expressions for twist-tree structure functions of Drell-Yan process}

By substituting Eqs.(\ref{eq:phi}), (\ref{eq:phiAalpha}) into Eq.~(\ref{htensor}), and then contracting
the lepton tensor and the hadronic tensor, one can get the expression of the cross section for the
Drell-Yan process up to order $1/Q$.
As shown in Ref.~\cite{Arnold2009}, there are forty-eight independent structure functions
for Drell-Yan processes from the collision of two polarized spin-$1/2$ hadrons beams, and twenty-four of them
are leading-twist observables.
Here we restrict ourself to consider the contributions at order $1/Q$.
We found that the angular distribution of the differential cross section of polarized
Drell-Yan process at twist-three level can be expressed as
\begin{align}
 \frac{d\sigma^{\text{twist-3}}}{d x_1 dx_2 d^2 \bm{q}_T \, d \Omega}
=&{ \alpha_{em}^2\over 3Q^2}  \sin 2\theta
 \Big \{
 \cos \phi \, F_{UU}^{\cos \phi}
 \; + \, S_{1L}
    \sin \phi \, F_{LU}^{\sin \phi}
 \; + \, S_{2L}
    \sin \phi \, F_{UL}^{\sin \phi}
+
 \, S_{1L} \, S_{2L}
   \cos \phi \, F_{LL}^{\cos \phi}
\nonumber \displaybreak[0] \\
& + \, |\vec{S}_{1T}| \Big[ \sin(\phi_1+\phi)
 F_{TU}^{\sin(\phi_{S_1}+\phi)}
 + \sin(\phi_{S_1}-\phi)
 F_{TU}^{\sin(\phi_{S_1}-\phi)}
 \Big]
\nonumber \displaybreak[0] \\
& + \, |\vec{S}_{2T}| \Big[ \sin(\phi_{S_2}+\phi)
 F_{UT}^{\sin(\phi_{S_2}+\phi)}
 + \sin(\phi_{S_2}-\phi)
 F_{UT}^{\sin(\phi_{S_2}-\phi)}
 \Big]
\nonumber \displaybreak[0] \\
& + \, S_{1L} \, |\vec{S}_{2T}| \Big[
 +  \cos (\phi_{S_2}+\phi) \, F_{LT}^{\cos (\phi_{S_2}+\phi)}
+\cos (\phi_{S_2}-\phi) \, F_{LT}^{\cos (\phi_{S_2}-\phi)}\Big]
\nonumber \\
& + \, S_{2L} \, |\vec{S}_{1T}| \Big[
 +  \cos (\phi_{S_1}+\phi) \, F_{TL}^{\cos (\phi_{S_1}+\phi)}
+\cos (\phi_{S_1}-\phi) \, F_{TL}^{\cos (\phi_{S_1}-\phi)}\Big]
\nonumber \\
&  + \, |\vec{S}_{1T}| \, |\vec{S}_{2T}|  \Big[
   \cos (\phi_{S_1} + \phi_{S_2}+\phi) F_{TT}^{ \cos (\phi_{S_1} + \phi_{S_2}+\phi)}
 + \cos (\phi_{S_1} + \phi_{S_2}-\phi) F_{TT}^{ \cos (\phi_{S_1} + \phi_{S_2}-\phi)}
\nonumber \\
&
 +\cos (\phi_{S_1} - \phi_{S_2}+\phi) F_{TT}^{ \cos (\phi_{S_1} - \phi_{S_2}+\phi)}+
\cos (\phi_{S_1} - \phi_{S_2} - \phi) F_{TT}^{ \cos (\phi_{S_1} - \phi_{S_2}-\phi)}\Big] \Big\}\,.
\label{eq:dy_angular}
\end{align}
We obtain sixteen twist-three structure functions, the angular dependences of which
are consistent with the results given in Eq.~(57) of Ref.~\cite{Arnold2009}.

To shorten the notation we will use following combinations,
since they always appear in the same way:
\begin{eqnarray}
\widehat{f} = x_1\left((1-c)\,f+c\,\tilde{f}\right),\,\, \widehat{\overline{f}}=
x_2\left(c\, \overline{f}+(1-c)\,\tilde{\overline{f}}\right)
\end{eqnarray}
where $f$ and $\tilde{f}$ stand for the twist-three quark distributions,
and $\overline f$ and $\tilde{\overline{f}}$  for the antiquark distributions, respectively.
Using the EOM relations in Eqs.(\ref{eq:fperptilde}) to (\ref{eq:htilde}), the structure functions for polarized Drell-Yan processes at twist-three thus are expressed as
\begin{align}
F_{UU}^{\cos\phi}  = &
 {2\over Q}\,\mathcal{C}\left[(\bm{h} \cdot \bm{k}_{1T})
\left (\widehat f^\perp \overline{f}_1
-{M_2\over M_1} h_1^\perp
\widehat{\overline {h}} \right) \,
- \,(\bm{h} \cdot \bm{k}_{2T}) \left (f_1\,\widehat{\overline f}\pp
-{M_1\over M_2} \widehat h \, \overline{h}\pp_1  \right)
\right]
\displaybreak[0] \label{eq:fuu}\\
F_{UL}^{\sin\phi}   = & - {2\over Q} \,\mathcal{C}\left[(\bm{h} \cdot \bm{k}_{1T})
\left
( \widehat g^\perp \,\overline{g}_{1L}
+ {M_2\over M_1}h_1^\perp \, \widehat{\overline h}_L \right)
- (\bm{h} \cdot \bm{k}_{2T})\left(f_1 \widehat{\overline f}\pp_L
+ {M_1\over M_2}\,\widehat h \,h_{1L}^\perp\right)\right]
\displaybreak[0]\label{eq:ful}\\
F_{LU}^{\sin\phi} = & {2\over Q}\,\mathcal{C}\left[
(\bm{h} \cdot \bm{k}_{1T})\left(  \widehat f_L^\perp \overline{f}_1
+ {M_2\over M_1}\,h_{1L}^\perp\, \widehat{\overline h} \right)
- (\bm{h} \cdot \bm{k}_{2T})
\left( g_{1L} \widehat{\overline g}\pp
+ {M_1\over M_2}  \widehat h_L \,\overline{h}_1^\perp\right)\right]
\displaybreak[0] \label{eq:flu}\\
 F_{UT}^{\sin(\phi_{S_2}-\phi)}
= &  {1\over Q}
\,\mathcal{C}\left[
 2M_2  f_1 \,\widehat{\overline f}_T +2 M_1 \,\widehat h\,h_1
+ \left(\bm{k}_{1T}\cdot \bm{k}_{2T}\right)
\left(  { \widehat f^\perp \,\overline{f}_{1T}\over M_2}
-  {\widehat g^\perp  \, \overline{g}_{1T} \over M_2}
- {h_{1}^\perp\,\widehat{\overline h}_T  \over M_1}
+ {h_{1}^\perp\, \widehat{\overline h}\pp_T \over M_1}
\right) \right]
\displaybreak[0]\label{eq:fut1} \\
 F_{UT}^{\sin(\phi_{S_2}+\phi)}
= &  {1\over Q}
\mathcal{C}\,\left[
- \left(2\left(\hat{\bm{h}}\cdot \bm{k}_{2T}\right)^2
- \bm{k}_{2T}^2\right)
\left( {f_1\, \widehat{\overline f}\pp_T \over M_2}
+ { M_1\,\widehat h \,h_{1T}^\perp\over M_2^2}  \right)\right.\nonumber\\
& \left. + \left(2\hat{\bm{h}}\cdot \bm k_{1T}\hat{\bm{h}}\cdot \bm k_{2T}
-\bm k_{1T}\cdot \bm k_{2T}\right)
\left(  { \widehat f^\perp \overline{f}\pp_{1T}\over M_2}
+  { \widehat g^\perp \, \overline{g}_{1T} \over M_2}
+ {h_{1}^\perp\, \widehat{\overline h}\pp_T  \over M_1}
+ {h_{1}^\perp\,\widehat{\overline h}_T \over M_1}
\right) \right]
\displaybreak[0] \label{eq:fut2}\\
 F_{TU}^{\sin(\phi_{S_1}-\phi)}
= &  {1\over Q}\,\mathcal{C}\left[
2 M_1 \,\widehat{f}_T \overline f_1
+ 2 M_2 \,\, h_1 \widehat{\overline h}
+ \left(\bm{k}_{1T}\cdot \bm{k}_{2T}\right)
\left(  { f_{1T}^\perp \widehat{\overline f}\pp\over M_1}
- { {g}_{1T}\widehat{\overline g}\pp
 \over M_1}
-  {\widehat{ h}_T \overline h\pp_{1} \over M_2}
+ \, {\widehat{h}\pp_T \overline h\pp_{1}\over M_2}
\right) \right]
\displaybreak[0] \\
F_{TU}^{\sin(\phi_{S_1}+\phi)}
= &  {1\over Q}\,\mathcal{C}\left[
- \left(2\left(\hat{\bm{h}}\cdot \bm{k}_{1T}\right)^2
- \bm{k}_{1T}^2\right)\left(
{\widehat{f}\pp_T \overline f_1 \over M_1}
+ { M_1 \,h_{1T}^\perp \widehat{\overline h}\over M_1^2} \right)\right.\nonumber\\
& \left.+\left(2\hat{\bm{h}}\cdot \bm{k}_{1T}\hat{\bm{h}}\cdot \bm{k}_{2T}
- \bm{k}_{1T}\cdot \bm{k}_{2T}\right)
\left({f_{1T}^\perp\widehat{\overline f}\pp \over M_1}
+ {g_{1T}\widehat{\overline g}\pp \,\over M_1}
+  {\widehat{h}_T \overline h\pp_{1}\over M_2}
+ {\,\widehat{h}\pp_T \overline h\pp_{1}\over M_2}
\right) \right]
\displaybreak[0] \\
F_{LL}^{\cos\phi}  = & - {2\over Q}\,\mathcal{C}\left[(\bm{h} \cdot \bm{k}_{1T})
\left (\widehat g_L^\perp \overline{g}_{1L}
\,+\,{M_2\over M_1}h_{1L}^\perp \,\widehat{\overline h}_L\right)
- (\bm{h} \cdot \bm{k}_{2T}) \left (g_{1L} \widehat{\overline g}\pp_L
+{M_1\over M_2} \,\widehat h_L \overline{h}_{1L}^\perp  \right)
\right]
\displaybreak[0] \\
F_{LT}^{\cos(\phi_2-\phi)} = &
 {1\over Q}\,\mathcal{C}
\left[  2M_2\, g_{1L} \widehat{\overline g}_T
+2M_1 \widehat h_L \,\overline{h}_{1}
-(\bm{k}_{1T}\cdot \bm{k}_{2T})
\left( {\widehat g_L^\perp \overline{g}_{1T}\over M_2}
+ {h_{1L}^\perp\,\widehat{\overline h}_T \over M_1}
+ {\widehat f_L^\perp \overline{f}_{1T}^\perp\over M_2}
- {h_{1L}^\perp \,\widehat{\overline h}\pp_T\over M_1}\right)
\right]
\displaybreak[0] \\
F_{LT}^{\cos(\phi_2+\phi)}   =&
{1\over Q}\mathcal{C}
\left[  \left(2\left(\hat{\bm{h}}\cdot \bm{k}_{2T}\right)^2
- \bm{k}_{1T}^2\right)
\left({g_{1L} \widehat{\overline g}\pp_T\over M_2}
+ { M_1\,\widehat h_L\,\overline h\pp_{1T}\over M_2^2}  \right)\right.\nonumber\\
& \left. -\left(2\hat{\bm{h}}\cdot \bm{k}_{1T}\hat{\bm{h}}\cdot \bm{k}_{2T}-\bm{k}_{1T}\cdot \bm{k}_{2T}\right)
\left(  {\widehat g_L^\perp \overline{g}_{1T}\over M_2}
+ {h_{1L}^\perp\,\widehat{\overline h}_T \over M_1}
- {\widehat f_L^\perp \overline{f}\pp_{1T} \over M_2}
+ {h_{1L}^\perp  \widehat{\overline h}\pp_T \over M_1}\right) \right]
\displaybreak[0] \\
 F_{TL}^{\cos(\phi_{S_1}-\phi)}
= & {1\over Q}\,\mathcal{C}
\left[- 2 M_1 \, \widehat g_T \, \overline{g}_{1L}
-2M_2\,h_{1}\widehat{\overline h}_L
+ (\bm{k}_{1T}\cdot \bm{k}_{2T})
\left( {g_{1T} \,\widehat{\overline g}\pp_L\over M_1}
+ { \widehat h_T \,\overline{h}\pp_{1L}\over M_2}
+ {f_{1T}^\perp\, \widehat{\overline f}\pp_L \over M_1}
- { \widehat h_T^\perp \,\overline{h}\pp_{1L}\over M_2}\right)
 \right]
\displaybreak[0] \\
F_{TL}^{\cos(\phi_{S_1}+\phi)}
= & {1\over Q}\mathcal{C}\left[
-  \left(2\left(\hat{\bm{h}}\cdot \bm{k}_{1T}\right)^2
- \bm{k}_{1T}^2\right) \left({ \widehat g_T^\perp \,\overline{g}_{1L}\over M_1}
+ { M_2 \,h_{1T}^\perp\, \widehat{\overline h}_L \over M_1^2}\right)\right.\nonumber\\
& \left.+ \left(2\hat{\bm{h}}\cdot \bm k_{1T}\hat{\bm{h}}\cdot \bm k_{2T}
- \bm k_{1T} \cdot \bm k_{2T}\right)
\left( {g_{1T} \,\widehat{\overline g}\pp_L \over M_1}
+ { \widehat h_T\,\overline{h}\pp_{1L}\over M_2}
- {f_{1T}^\perp \,\widehat{\overline f}\pp_L \over M_1}
+ { \widehat h_T^\perp \,\overline{h}\pp_{1L}\over M_2}\right) \right]
\displaybreak[0] \\
F_{TT}^{\cos(\phi_a+\phi_b-\phi)}
=&{1\over Q}\mathcal{C}\left[
 (\bm h \cdot \bm k_{1T})\,
\left( {M_2\,f_{1T}^\perp \,\widehat{\overline f}_T \over M_1}
- {M_2 \,g_{1T} \,\widehat{\overline g}_T  \over M_1}
-\widehat h_T \overline h_1
- \widehat{h}\pp_T \overline{h}_1
\right)\right.\nonumber\\
&-\left.
(\bm h \cdot \bm k_{2T})\,
\left( {M_1 \widehat g_T \,\overline{g}_{1T}\over M_2}
- {M_1\,\widehat f_T \overline f\pp_{1T} \over M_2}
+ h_1\, \widehat{\overline h}_T
+ h_1 \,\widehat{\overline h}\pp_T \right)\right]
\displaybreak[0] \\
F_{TT}^{\cos(\phi_a+\phi_b+\phi)}
= & {1\over Q}\,\mathcal{C}
\left[\left(4(\hat{\bm{h}}\cdot \bm{k}_{2T})(\hat{\bm{h}}\cdot \bm{k}_{1T})^2
- 2(\hat{\bm{h}}\cdot \bm{k}_{1T})
(\bm{k}_{1T}\cdot \bm{k}_{2T})
- (\hat{\bm{h}}\cdot \bm{k}_{2T})\bm{k}_{1T}^2\right)
\right.\nonumber\\
&\times \left(
{\widehat f\pp_T \,\overline f\pp_{1T} \over 2M_1 M_2}
-{ \widehat g_T^\perp \,\overline{g}_{1T} \over 2M_1 M_2}
- {h_{1T}^\perp  \,\widehat{\overline h}_T \over 2M_1^2 }
- {h_{1T}^\perp \,\widehat{\overline h}\pp_T \over 2M_1^2 } \right)\nonumber\\
&-\left(4(\hat{\bm{h}}\cdot \bm{k}_{1T})(\hat{\bm{h}}\cdot \bm{k}_{2T})^2 -2(\hat{\bm{h}}\cdot \bm{k}_{2T}) (\bm{k}_{1T}\cdot \bm{k}_{2T})-(\hat{\bm{h}}\cdot \bm{k}_{1T})\bm{k}_{2T}^2\right)
\nonumber\\
&\times \left.\left( {\,f_{1T}^\perp\, \widehat{\overline f}\pp_T \over 2M_1 M_2}
- { g_{1T}\,\widehat{\overline g}\pp_T  \over 2M_1 M_2}
- { \widehat{h}_T \,\overline h\pp_{1T} \over 2M_2^2 }
- {\widehat h\pp_T \,\overline{h}\pp_{1T} \over 2M_2^2 }
\right)\right]
\displaybreak[0] \\
F_{TT}^{\cos(\phi_a-\phi_b+\phi)}
=&-{1\over Q}\,\mathcal{C}\left[
(\bm h \cdot \bm k_{1T})\,
\left( {M_2\,f_{1T}^\perp \, \widehat{\overline f}_T\over M_1}
- {M_2 {g}_{1T}\widehat{\overline g}_T \over M_1}
- \widehat h_T \overline h_1
- \widehat{h}\pp_T   \, \overline{h}_1
\right)\right.\nonumber\\
& - \left(2(\hat{\bm{h}}\cdot \bm{k}_{1T})
(\bm{k}_{1T}\cdot \bm{k}_{2T})
- (\hat{\bm{h}}\cdot \bm{k}_{2T})\bm{k}_{1T}^2\right)\nonumber\\
&\times\left. \left(
{\widehat{f}\pp_{T} \overline f\pp_{1T}  \over 2M_1 M_2}
+{\widehat g_T^\perp  \,\overline{g}_{1T} \over 2M_1 M_2}
+{h_{1T}^\perp \,  \widehat{\overline h}_T \over 2M_1^2}
-{ h_{1T}^\perp\widehat{\overline h}\pp_T\over 2M_1^2 }
\right)\right]
\displaybreak[0] \\
F_{TT}^{\cos(\phi_a-\phi_b-\phi)}
=&{1\over Q}\,\mathcal{C}\left[
(\bm h \cdot \bm k_{2T})\,\left(
{M_1  \widehat{f}_T\,\overline f\pp_{1T} \,\over M_2}
-{M_1 \widehat g_T \,\overline{g}_{1T}\over M_2}
- h_1 \,\widehat{\overline h}_T
- h_1 \widehat{\overline h}\pp_T\right)\right.\\
&- \left(2(\hat{\bm{h}}\cdot \bm{k}_{2T}) (\bm{k}_{1T}\cdot \bm{k}_{2T})
-(\hat{\bm{h}}\cdot \bm{k}_{1T})\bm{k}_{2T}^2\right)\nonumber\\
&\times\left.\left( {f_{1T}^\perp\, \widehat{\overline f}\pp_{T} \over 2M_1 M_2}
+ { g_{1T}\widehat{\overline g}\pp_T  \over 2M_1 M_2}
+ {\widehat{h}_T    \overline h\pp_{1T} \over 2M_2^2}
- {\widehat h_T^\perp \overline{h}\pp_{1T}\over 2M_2^2 }
\right)\right]\label{eq:ftt}
\end{align}
In the above equations we have used the notation:
\begin{equation}
\begin{split}
\mathcal{C}\big[w(\bm{k}_{1T},\bm{k}_{2T})f\bar{g} \big] =&  \sum_q e_q^2 \int d^2 \bm{k}_{1T} ~ d^2 \bm{k}_{2T} ~ \\
&\times \delta^{(2)}\left(\bm{q}_T - \bm{k}_{1T} - \bm{k}_{2T} \right)  w(\bm{k}_{1T},\bm{k}_{2T}) \\
 & \times \left[ f^{q}( x_1 ,k_{1T}^2) g^{\bar{q}}( x_2 , k_{2T}^2)\right.
 \\
 &\left.+ f^{\bar{q}}( x_1 ,k_{1T}^2) g^q( x_2 , k_{2T}^2)\right].
\end{split}
\end{equation}
Also, we have applied the following combinations for certain distribution functions:
\begin{align}
f_T^{}(x, k_T^2) &=
  f_T'(x, k_T^2) - \frac{k_T^2}{2 M^2}\, f_T^{\perp}(x, k_T^2),
\\
g_T^{}(x, k_T^2) &=
  g_T'(x, k_T^2) - \frac{k_T^2}{2 M^2}\, g_T^{\perp}(x, k_T^2),
\\
\tilde f_T^{}(x, k_T^2) &=
  \tilde f_T'(x, k_T^2) - \frac{k_T^2}{2 M^2}\, \tilde f_T^{\perp}(x, k_T^2),
\\
\tilde g_T^{}(x, k_T^2) &=
  \tilde g_T'(x, k_T^2) - \frac{k_T^2}{2 M^2}\, \tilde g_T^{\perp}(x, k_T^2),
\\
h_{1}(x, k_T^2) &=
  h_{1T}(x, k_T^2) - \frac{k_T^2}{2 M^2}\, h_{1T}^{\perp}(x, k_T^2),
\end{align}

Equations (\ref{eq:fuu}) to (\ref{eq:ftt}) represent a main result of this work.
They are complementary to the leading-twist structure functions of Drell-Yan processes.
We note that, after being integrated over $\bm q_T$, the expression in
Eq.~(\ref{eq:dy_angular}) reduces to the following result~\cite{Tangerman:1994bb,Boer:1997bw}:
\begin{align}
 \frac{d\sigma^{\text{twist-3}}}{d x_1 dx_2 \, d \Omega}
=&{ \alpha_{em}^2\over 3Q^2}  \sin 2\theta \,{2\over Q} \Big\{
\lvert S_{1T}\rvert \sin(\phi_{S_1}-\phi)
\left( M_1 \widehat f_T\,\overline f_1
 + M_2 h_1\, \widehat{\overline h} \right)
\nonumber\\
&
+\lvert S_{2T}\rvert \sin(\phi_{S_2}-\phi)
\left( M_2\,f_{1}\,\widehat{\overline f}_T
+ M_1 \widehat h\,\overline h_1 \right)\nonumber\\
&+ S_{1L} \lvert S_{2T}\rvert\cos(\phi_{S_2}-\phi)
\left( M_2\, g_1 \widehat{\overline g}_T
+ M_1 \widehat h_L \,\overline{h}_{1}\right)
\nonumber\\
&- S_{2L} \lvert S_{1T}\rvert \cos(\phi_{S_1}-\phi)
\left( M_1\, \widehat g_T  \overline g_1
+M_2 h_1 \widehat{\overline h}_L\right)\Big\}.\label{eq:qtint}
\end{align}
where the distributions are the transverse-momentum integrated version.
The last two terms in the curly brackets were given in Ref.~\cite{Tangerman:1994bb}, while the
first two terms were given in Ref.~\cite{Boer:1997bw}.
However, as $f_T(x)$ and $h(x)$ vanish because of time reversal invariance~\cite{Goeke:2005hb},
the first two terms in Eq.~(\ref{eq:qtint}) only receive contributions from the interaction-dependent
twist-three distributions $\tilde{f}_T$ and $\tilde{h}$.
The function $\tilde{f}_T(x)$ is related~\cite{Boer:2003cm,Ma:2003ut} to the collinear twist-3 Efremov-Teryaev-Qiu-Sterman
(ETQS) function $T_F(x,x)$~\cite{Efremov:1981sh,Efremov:1984ip,Qiu:1991pp}.
The role of the ETQS function in the single spin asymmetry $A_N$ in Drell-Yan were studied intensively in the
literature~\cite{Hammon:1996pw,Boer:1999si,Boer:2001tx,Ma:2003ut},
and has been revisited in Refs.~\cite{Anikin:2010wz} and
\cite{Zhou:2010ui} recently.
It was found that different theoretical approaches lead to different results.
For example, an additional factor $2$ was found in Ref.~\cite{Anikin:2010wz} while
factor $1/2$ was gained in Ref.~\cite{Zhou:2010ui}, in comparison with the expression
for $A_N$ in Refs.~\cite{Boer:1997bw,Boer:1999si,Boer:2001tx}.
We note that the normalization of the asymmetry of our result (in the Collins-Soper frame) agrees with that in
Refs.~\cite{Boer:1997bw,Boer:1999si,Boer:2001tx}, but not with that in
Refs.~\cite{Anikin:2010wz,Zhou:2010ui}.

In the following we present further comments on our result.
\begin{itemize}

\item Certain structure functions in Eqs.(\ref{eq:fuu}) to (\ref{eq:ftt}) satisfy the following symmetric or antisymmetric property:
\begin{align}
&F_{UL}^{\sin\phi}=-F_{LU}^{\sin\phi},~~
F_{UT}^{\sin(\phi_{S_2}-\phi)} = F_{TU}^{\sin(\phi_{S_1}-\phi)},\\
&F_{UT}^{\sin(\phi_{S_2}+\phi)} = F_{TU}^{\sin(\phi_{S_1}+\phi)},~~
F_{LT}^{\cos(\phi_{S_2}-\phi)} =-F_{TL}^{\cos(\phi_{S_1}-\phi)},~~
F_{LT}^{\cos(\phi_{S_2}+\phi)} =-F_{TL}^{\cos(\phi_{S_1}+\phi)},
\end{align}
which agree with the general analysis given in Ref.~\cite{Arnold2009}.

\item The expressions of the twist-three structure functions depend on the choice of
the dilepton rest frame, characterized by the parameter $c$.
This feature is different from that of leading-twist structure functions, whose expressions
of which are the same in different dilepton rest frames.

\item  Among the 16 twist-three distributions, 12 of them appear in
the twist-three structure functions for Drell-Yan processes, including the new distributions $g^\perp$~\cite{Bacchetta:2004zf} and $f_T^\perp$~\cite{Goeke:2005hb}.
The first one appears in both the single longitudinally and transverse polarized
Drell-Yan processes, while the later one appears in single and double transversely
polarized Drell-Yan processes.
These new twist-three TMDs are quite interesting, since evidence of their existence will indicate the
necessity of introducing the gauge-link direction in the decomposition of the correlator~\cite{Bacchetta:2004zf,Goeke:2005hb}.
They represent new nucleon parton structure information that has not been explored before.
Calculations~\cite{Afanasev:2003ze,Metz:2004je} in a diquark model show that the contribution
from $g^\perp$ should be included in order to give a complete description of the asymmetries
in longitudinally polarized SIDIS.

The distributions $e$, $e_L$, $e_T$ and $e_T^\perp$ do not contribute here,
since we have not considered the lepton polarization.
These four distributions could be studied in SIDIS with a polarized lepton
beam~\cite{Bacchetta:2006tn,Avakian:2003pk,Yuan:2003gu,Aghasyan:2011ha}.

\item Equation (\ref{eq:fuu}) shows that the combination $\widehat f^\perp \otimes \overline f_1$
or $h_1^\perp\otimes \widehat{\overline h}$ and so on can lead to a $\cos\phi$ asymmetry in
unpolarized Drell-Yan processes, which is similar to the case in SIDIS~\cite{Levelt:1994np}.
 Apart from the well-known $\cos2\phi$ distributions,
the measurements of dilepton production in unpolarized hadron-hadron collision
also show a $\cos\phi$ angular dependence~\cite{na10a,na10b,conway}.
It has been shown that the QCD corrections can generate such angular distribution~\cite{Collins:1978yt,Cleymans:1978ip,Lindfors:1979rc,mo94,bv06,Berger:2007jw}.
Our study shows that there is an alternative mechanism for the $\cos\phi$ asymmetry in the unpolarized Drell-Yan process, due to the presence of twist-three TMD distributions.

\item In the region where the TMD framework is assumed to be valid, there is a suppression
factor $q_T/Q$ for twist-three structure functions.
Therefore the asymmetries arising from twist-three structure functions are supposed to be smaller
than the leading-twist observables.
However, the asymmetries are also directly determined by the size of the twist-three TMD
distributions.
It is not known whether there are positivity bounds to constrain  twist-three TMD distributions,
like the case of leading-twist distributions~\cite{Bacchetta2000}.
Sizable twist-three TMD distributions could lead to nonvanishing asymmetries in Drell-Yan process.
Furthermore, the SIDIS measurements on fixed targets~\cite{
Aubert:1983cz,Arneodo:1986cf,Airapetian:1999tv,Airapetian:2001eg,Avakian:2003pk,
Mkrtchyan:2007sr,Kafer:2008ud,Aghasyan:2011ha} show that the twist-three asymmetries are not small. This also encourages the corresponding measurements of twist-three effects
in Drell-Yan processes, especially for fixed-target experiments~\cite{PAX_05,COMPASSpro,Goto:2010zz,Goto:2011zz,Liu:2010kb}.

\item For each structure function, there are several combinations of twist-two distribution and
twist-three distribution which can contribute, similar to the twist-three results of
SIDIS process.
This makes twist-three parton distributions more difficult to be probed in high energy processes
than the twist-two ones.
Further theoretical and experimental studies are needed to provide more constrains
the size of different TMD twist-three distributions.

\item  We point out that our calculation is based on a generalization of the TMD factorization to the twist-three level. Therefore the correctness of our results relies on the validation
of the twist-three TMD factorization.
Unlike the twist-three collinear factorization, which has been widely applied in
SIDIS and Drell-Yan, the TMD factorization formalism at twist-three level has not been
established yet.
The main challenge for twist-three TMD observables is that when one calculates
the twist-three TMDs, there are light-cone divergences~\cite{Gamberg:2006ru}
for which it has not been understood how to control them at order $1/Q$.
This does not necessarily means that twist-three TMD factorization cannot be developed.
Further study is needed to overcome this difficulty.

\end{itemize}

\section{conclusion}

Drell-Yan process has been recognized as an important tool to study the
structure of the nucleon. In this work, we have studied  polarized Drell-Yan processes from the collision
of two polarized spin-$1/2$ hadrons beams at order $1/Q$, based on
the framework of TMD factorization.
We find that, among a total of twenty-four subleading-twist structure functions in Drell-Yan process,
sixteen of them are at twist-three level and can be expressed as combinations of
twist-two and twist-three TMD distribution functions.
We give the complete expressions for these structure functions, for each of which there are
several twist-three distributions that contribute.
Based on our result, we point out that twist-three distributions can provide an alternative
explanation for the $\cos\phi$ angular dependence in unpolarized Drell-Yan processes.
The measurements of the asymmetries at order $1/Q$ in Drell-Yan process therefore
can provide  useful information on the twist-three TMD distributions and the
multi-parton correlations in the nucleon.

{\bf Acknowledgements} This work is supported by FONDECYT (Chile) Projects No.~11090085,
No.~1100715, by Project Basal FB0821,
by NSFC (China) Project No.~11005018, and by Teaching and Research Foundation for 
Outstanding Young Faculty of Southeast University.

\end{document}